\documentclass[acmsmall,nonacm]{acmart}

\AtBeginDocument{%
  }

\setcopyright{acmlicensed}
\copyrightyear{2018}
\acmYear{2018}
\acmDOI{XXXXXXX.XXXXXXX}

\acmConference[Conference acronym 'XX]{Make sure to enter the correct
  conference title from your rights confirmation email}{June 03--05,
  2018}{Woodstock, NY}
  
\acmISBN{978-1-4503-XXXX-X/2018/06}

\definecolor{violet}{RGB}{120, 100, 230}
\definecolor{green}{RGB}{0, 135, 71}
\definecolor{black}{RGB}{0, 0, 0}
\definecolor{yellow}{RGB}{200, 180, 0}
\definecolor{bluegray}{RGB}{54, 69, 79}
\definecolor{nightBlue}{RGB}{45,78,159}

\usepackage{booktabs,multirow,tabularx,array, graphicx,subcaption}

\begin{document}

\title{Can Vibe Coding Produce Accessible Websites: An Empirical Study on LLM Coding Agents and Skills}

\author{Syed Fatiul Huq}
\email{fsyedhuq@uci.edu}
\orcid{0000-0002-9039-5848}
\affiliation{%
  \institution{University of California, Irvine}
  \city{Irvine}
  \state{California}
  \country{USA}
}

\author{Md. Asif Haider}
\email{asifh1@uci.edu}
\orcid{0000-0002-8850-657X}
\affiliation{%
  \institution{University of California, Irvine}
  \city{Irvine}
  \state{California}
  \country{USA}
}

\author{Abu Jafar Saifullah}
\email{bsse1109@iit.du.ac.bd}
\orcid{0009-0005-8334-6019}
\affiliation{%
  \institution{University of Dhaka}
  \city{Dhaka}
  \country{Bangladesh}
}

\author{Thomas Zimmermann}
\email{tzimmer@uci.edu}
\orcid{0000-0003-4905-1469}
\affiliation{%
  \institution{University of California, Irvine}
  \city{Irvine}
  \state{California}
  \country{USA}
}

\author{Sam Malek}
\email{malek@uci.edu}
\affiliation{%
  \institution{University of California, Irvine}
  \city{Irvine}
  \state{California}
  \country{USA}
}

\renewcommand{\shortauthors}{Huq et al.}

\begin{abstract}
  Large language model (LLM) coding agents increasingly allow developers and non-professionals to create websites through natural-language instructions, with limited inspection of the generated source code. Although this ``vibe coding'' workflow lowers barriers to software creation, it may also propagate accessibility defects at scale. We investigate the accessibility of websites generated end-to-end by contemporary coding agents and whether agent-level accessibility interventions improve their output. We generated 90 websites for three projects using combinations of three coding agents, four language models, and multiple reasoning-effort levels. We evaluated complete interaction journeys using axe-core, IBM Equal Access Accessibility Checker, and QualWeb,  and converged reports to observe violation prevalence and violated WCAG Success Criterion (SC). None of the 90 baseline websites was fully accessible, with an average of 80.43 violated interface elements. 13 WCAG SCs were violated at least once across the runs. Color contrast was the most prevalent issue, followed by labeling, semantic-structure, keyboard, and visualization failures. We reran 24 matched configurations with a general accessibility Skill and a contrast-specific Skill. Both significantly improved accessibility scores, but neither eliminated violations. User testing with blind individuals corroborated the experiential effects of the detected inaccessibilities. Moreover, functional correctness decreased with the inclusion of Skills. Our findings show that accessibility does not reliably emerge from vibe coding, and agentic interventions can both improve some accessibility issues and introduce new functional ones.
\end{abstract}



\maketitle

\section{Introduction}
The advancement of Large Language Models (LLMs) in the software engineering domain has ushered in a new paradigm of software development: vibe coding. Introduced in 2024~\footnote{https://x.com/karpathy/status/1886192184808149383}, the term ``vibe coding'' refers to, in public lexicon as well as academic corpora~\cite{chou2026building}, the process of developing software merely via natural language prompts. The tool support for vibe coding has evolved from chat interfaces and IDE autocomplete features, to full-fledged AI-focused development environments. This paradigm shift has been adopted by seasoned developers for easy task delegation and an increase in productivity~\cite{mohamed2026impact}. It has also become a boon to non-professionals, as vibe coding arguably reduces the necessity for learning, comprehending and manipulating source code. Its promise of democratizing software development, however, runs in parallel with a decrease in oversight. That is especially problematic when AI-generated code has been shown to underperform in non-functional aspects like security~\cite{zhao2025vibe}, coding conventions and API constraints~\cite{zhong2025swe}, and more relevant for this paper, accessibility. When human developers are only evaluating their application for aesthetics and functional correctness, the software space may soon be littered with vulnerable, incongruent and inaccessible applications.

Accessibility in software refers to how well people with different disabilities, who constitute 16\% of the global population~\cite{WHO}, can navigate and interact with web, mobile and Desktop applications. This requires a careful design and implementation of User Interface (UI) elements. For instance, if a developer does not write appropriate labels, constructs such as images, icon buttons and input forms become meaningless and inaccessible to blind individuals. Regulations such as the Americans with Disabilities Act (ADA)~\cite{ADA} and The European Accessibility Act (EAA)~\cite{EAA} now legally require accessibility compliance for any published website, following Web Content Accessibility Guidelines (WCAG) 2.2~\cite{wcag2024web} as the reference documentation. To help developers understand, detect and fix accessibility issues, several tools~\cite{WebAIM2023WAVE, axecore, ibmequalaccess, qualweb} and techniques~\cite{jain2025automated, durgam2023dynamic, Duarte, chiou2021detecting, chiou2023detecting, chiou2023bagel, bajammal2021semantic} have been developed, which automatically audit the code or website. Recently, LLMs have been used to test previously undetectable issues, owing to its multi-modality and subjectivity comprehension~\cite{he2025enhancing, he2026reflair, huq2025automated}. The results show potential of LLMs as an utility for accessible development.
Some prior studies have specifically experimented with LLM-generated code and accessibility~\cite{abu2025can, aljedaani2024does, mowar2025codea11y, leedy2025accessibility, aljedaani2026multi, suh2026empirical}. They showed that LLMs produce slightly more accessible code than humans. But the quality is still subpar, violating many accessibility criteria. For improvement, intentional and iterative prompting techniques were employed. But inaccessibility was never fully eradicated.

Three gaps remain from prior work. First, literature examines generated snippets or static pages rather than websites developed end-to-end by coding agents. Second, evaluations commonly audit an initial page, even though violations can appear only after navigation, form entry, state changes, or data visualization. Third, explicit accessibility prompts require the user to know that accessibility must be requested. This assumption is poorly matched to vibe coding, whose users may neither inspect the implementation nor know how to formulate accessibility requirements. If accessibility depends on every user remembering to ask for it, inaccessible output may remain the default.

We address these gaps through a two-part empirical study. Our first research question asks:

\begin{quote}
\textbf{RQ1: How accessible are websites generated by LLM coding agents?}
\end{quote}

To answer RQ1, we generated 90 websites across three projects: an educational course catalog, a meal-planning application, and a travel journal with interactive visualizations. The projects and their prompts were informed by prior work, WCAG coverage, and a mined dataset of 748 AI-assisted website repositories. We evaluated combinations of three coding agents (Codex~\cite{openaicodex2026}, Claude Code~\cite{claudecode2026}, and OpenCode~\cite{opencode2026}), four models (GPT~\cite{openaigpt2026}, Sonnet~\cite{anthropicsonnet2026}, DeepSeek~\cite{deepseekmodel2026}, and Qwen~\cite{qwenmodel2026}), and multiple reasoning-effort levels. Rather than auditing a single static page, we used Playwright~\cite{microsoft2026playwright} journeys to exercise each website's pages and functionality, and collected accessibility results at checkpoints throughout those journeys. We converged reports from axe-core~\cite{axecore}, IBM Equal Access Accessibility Checker~\cite{ibmequalaccess}, and QualWeb~\cite{qualweb} and calculated an \emph{accessibility score} from violation prevalence and the breadth of violated WCAG Success Criteria (SC). We conducted a small user study with blind users to complement automated findings. We additionally analyzed agent trajectories for development phases, reasoning summaries and verification actions.

Our second research question asks:

\begin{quote}
\textbf{RQ2: To what extent do accessibility-focused Skills change agents' behavior and the accessibility of the websites they generate?}
\end{quote}

To answer RQ2, we mined and characterized accessibility-related Skills~\cite{anthropic2026agentskills} published on GitHub, then selected two representative interventions: a general accessibility-audit Skill and a specialized Skill containing contrast-checking scripts. For each project, we selected configurations across accessibility-score quartiles while preserving representation of all agents and models. We reran 24 matched configurations with each Skill, producing 48 additional websites. This paired design allowed us to compare the generated output and the agent trajectory with and without an intervention, including accessibility-related reasoning, tool use, and verification activity.

Our results reveal a persistent gap between functional completion and accessible implementation. None of the 90 baseline websites was fully accessible. Their mean accessibility score was 88.33, with an average of 80.43 violated interface elements and 3.61 distinct violated Success Criteria. Color contrast failed in all 90 websites, while failures involving accessible names, information and relationships, label-in-name consistency, keyboard operation, form labeling, and graphical content appeared across projects. By contrast, 80 of the 90 websites received a perfect functional-correctness score, and functional correctness was not associated with accessibility. Model choice was significantly associated with accessibility score, but agent and reasoning-effort level were not. Although nearly half of trajectories with readable reasoning mentioned accessibility at least once, none of the baseline runs performed an accessibility-specific verification action. Evaluating websites with duplicate configurations revealed variability in accessibility, functionality and trajectories, proving the impact of LLM's non-determinism.

Accessibility Skills improved accessibility outcomes and trajectory behavior, but regressed functionally. The general Skill raised the mean score of 24 matched websites from 88.06 to 92.13, a significant mean increase of 4.07 points, while the contrast-specific Skill raised it to 92.56 and reduced mean contrast-violation prevalence from 12.53\% to 3.98\%. Skill-enabled runs devoted substantially more activity to verification and included explicit accessibility checks. Nevertheless, no Skill-enabled website became fully accessible. Furthermore, functional correctness fell, as general Skill produced a significant mean decrease of 19.01 points. These results show that accessibility interventions can be effective, but must be paired with well-scoped tooling, and regression checks that preserve both accessibility and requested functionality.

This paper makes the following contributions:

\begin{itemize}
    \item An end-to-end accessibility evaluation of 90 agent-generated websites across three coding agents, four models, and multiple reasoning-effort levels, using interaction journeys rather than isolated static pages.
    \item A matched evaluation of accessibility Skills, quantifying their effects on accessibility outcomes and agent trajectories.
    \item Implications for coding-agent design: accessibility guidance and executable testing should be supplied and invoked by default, followed by iterative, regression-aware remediation.
    \item A dataset of 4,138 GitHub repositories with accessibility Skills as a reference for further analysis of agentic interventions for accessible website generation.
\end{itemize}

The remainder of the paper is organized as follows. Section~\ref{sec:background} contains the literature review. Section~\ref{sec:method} describes our study design. Sections~\ref{sec:rq1} and \ref{sec:rq2} detail the results of the two research questions. Section~\ref{sec:discussion}, \ref{sec:threats} and \ref{sec:conclusion} conclude the paper with implications, future directions and threats to validity.

\section{Related Work}
\label{sec:background}

Since the emergence of LLM chatbots and coding assistants, examining the accessibility implications of LLM-generated code has been of interest. Doush et al.~\cite{abu2025can} generated HTML code for 11 specific web components, such as a combobox and a login form, using ChatGPT, CoPilot, Claude and Grok. The generated code was evaluated against author-written ``ideal'' code that followed accessibility guidelines. Aljedaani et al.~\cite{aljedaani2024does} 
evaluated websites written by developers with ChatGPT assistance, subsequently incorporating audit results into the prompt. The code was evaluated using WAVE~\cite{WebAIM2023WAVE} and IBM AChecker~\cite{ibmequalaccess}. Leedy et al.~\cite{leedy2025accessibility} prompted website builders like Wix and Framer with three levels of accessibility-aware prompts to generate three representative websites. All three studies found LLMs lacking in creating accessible HTML code.

Two studies aimed to develop interventions. Aljedaani et al.~\cite{aljedaani2026multi} expanded on their prior work to create a multi-agent system that audits the generated website for accessibility, fixes the issues and iterates until satisfied. Suh et al.~\cite{suh2026empirical} developer FeedA11y that also prompts the LLM with audit results for improved accessibility. Both demonstrate an increase in accessibility over the baseline.

These studies laid the foundation for evaluating LLMs for accessibility. Contemporary coding agents, like Codex~\cite{openaicodex2026} and Claude Code~\cite{claudecode2026}, which are built on reasoning models and integrated with tools, conduct end-to-end website building. This paradigm shift, where deliberate prompting and careful HTML parsing is not prioritized anymore~\cite{chou2026building}, leaves prior suggestions for prompt engineering no longer practical. Our work addresses this gap by evaluating the websites generated by these coding agents and incorporating agentic interventions in the form of Skills.
\section{Study Design}
\label{sec:method}

\subsection{Artifacts}
\label{subsec:artifacts}

The artifacts for this empirical study were websites generated by coding agents. We use the term \emph{website instance} to refer to an individual generated website, created using a configuration of coding agent, language model and reasoning effort level, under one of three projects.

\subsubsection{Projects}
\label{subsubsec:projects}

In an effort to maximize ecological representativeness, we triangulated with three conditions to determine the selection of projects and their corresponding prompt, assets and rules.

Firstly, we mined GitHub for existing websites created by or with the assistance of AI, characterized by traces of LLM contribution in the repository. Contribution was checked on commit messages and authorship (e.g., ``authored by Claude''), and the existence of agent-specific files (e.g., \texttt{CLAUDE.md}). From the 2,147 resulting repositories, 748 were selected as websites, determined by the existence of implementation-related files such as \texttt{index.html}, \texttt{package.json}, or by the involvement of GitHub Pages. Next, websites were categorized with a multi-agent labeler (Claude Sonnet-5) based on the live page, description and file tree. The top categories after manual inspection were web apps (194), business front (52), content (39), portfolio (31), game (20) and e-commerce (13).

Next, we followed the artifact selection by Thorgeirsson et al.~\cite{thorgeirsson2026computer} to narrow down the intent of the websites. They used the Delphi method~\cite{linstone1975delphi}, employing 8 experts to independently compile and anonymously judge ideas for websites to be vibe coded. Constraints included interdisciplinarity, real-world use case and GUI-first implementation. The resulting three representative websites were: a course platform, a meal planner and a visualization website.

Lastly, we analyzed WCAG SCs~\cite{wcag2024web} and ACT rules~\cite{w3cwai2026actrules} to find the web components most vulnerable to inaccessibility and covered by automated accessibility checkers. These would inform us about which features to include in the websites, so that we can observe LLM's treatment of all types of elements. The resulting elements include text, images, buttons, links, tables, form elements (e.g., text input, \texttt{checkboxes}, \texttt{dropdowns}, etc.), \texttt{iframes}, and \texttt{svg}.

Based on this process, we formulated three projects, Project 1 - Catalog, Project 2 - Planner, and Project 3 - Travel. We denote them by
\(\mathcal{P} = \{P_1, P_2, P_3\}\).
For each project \(P_p\), we define its fixed specification as
\[
P_p = (\pi_p, X_p, R_p),
\]
where \(\pi_p\) is the base prompt, \(X_p\) is a set of provided assets,
and \(R_p\) is a project-specific rule set. The specification is shared
by all website instances generated for \(P_p\). Their description and relevant data are presented in Table \ref{tab:method-artifacts-website}. 

\begin{table*}[t]
    \centering
    \scriptsize
    \renewcommand{\arraystretch}{1.2}
    \caption{Selected projects for website creation.}
    \label{tab:method-artifacts-website}
    \begin{tabularx}{\textwidth}{
        >{\raggedright\arraybackslash}p{0.1\textwidth}
        >{\raggedright\arraybackslash}X
        >{\raggedright\arraybackslash}p{0.12\textwidth}
        >{\raggedright\arraybackslash}X
        >{\centering\arraybackslash}p{0.05\textwidth}
        >{\centering\arraybackslash}p{0.05\textwidth}
        >{\centering\arraybackslash}p{0.05\textwidth}
    }
        \toprule
        \textbf{Project} &
        \textbf{Description} &
        \textbf{Category} &
        \textbf{Target Elements} &
        \textbf{Prompt Size} &
        \textbf{Assets Count} &
        \textbf{Rules Count} \\
        \midrule

        $P_1$ - Catalog &
        An educational platform showcasing their offered courses, syllabus and materials like documents and videos. &
        Business front - course platform &
        Navigation menus, course cards, tables, links, embedded media, and document viewers. &
        248 words &
        7 &
        18 \\

        $P_2$ - Planner &
        A meal prepping web app to create recipes, incorporate recipes in a weekly planner, and manage a shopping list for planned meals. &
        Web app - meal planner &
        Forms, checkboxes, comboboxes, listboxes, sliders, switches, text inputs, and calendar controls. &
        680 words &
        9 &
        19 \\

        $P_3$ - Viz &
        A website to visualize the itinerary for a trip and display its expenses with charts. &
        Content - travel journal &
        Interactive charts, tables, nested controls, and data visualizations. &
        225 words &
        3 &
        9 \\

        \bottomrule
    \end{tabularx}
\end{table*}

Base prompts $\pi_p$ included a list of expected features, based on the features of our mined websites in the same category and relevant topic. When formulating the exact prompt, the goal was to implicitly incorporate each project's target elements. The prompt would mention intended features (e.g., ``The right side of the course page should function as a persistent material-viewing area.''), without specifying implementation details (e.g., ``create an \texttt{iframe} component to display course material''). We conducted multiple iterations of site generation, code inspection and refinement with Codex and Claude Code, checking in each iteration whether all target elements were present. Prompts were finalized once that was accomplished. 

Other than project-specific base prompts, all projects were given restrictive prompts. The restriction enforced a focus on front-end implementation, which is necessary for accessibility evaluation, without developing server-side support. Other than prompting, we steered the agents with technology scaffolds, injecting React and Vite dependencies to a \texttt{package.json} file. The choice of React and Vite was influenced by our GitHub dataset.

We provided each website with a set of asset files ($X_p$) in \texttt{.csv} format containing relevant data. For instance, $P_3$-Travel contained three asset files: \textit{trip\_days}, \textit{itinerary} and \textit{expenses}. These listed information about the day-to-day plans and expenses of a fictitious trip across Europe. This controlled feature variability across runs and minimized effort from agents in constructing dummy data, which often redirected focus from website development to data creation. A consistent reference data also helped with formulating test cases for the automated evaluation.

Lastly, from the final prompt and assets, we mapped each project to a rule set $R_p$, containing a sequence of expected functionalities. For instance, the rule set for $P_2$-Planner specified actions for creating a new recipe with specific values for form fields (e.g., title, serving size, ingredients etc.), adding recipes to the meal planner, and checking off items from the shopping list.

Base prompts $\pi_p$, assets $X_p$ and rule sets $R_p$ are provided in the replication package \cite{replication_package}.

\subsubsection{Agents}
\label{subsubsec:agents}

A website instance under a project $P_p$ is generated by an agent with \emph{configuration} $c = (a,m,e)$, where coding agent $a \in \mathcal{A}$, language model $m \in \mathcal{M}$, and reasoning effort $e \in \mathcal{E}$.

For language models $\mathcal{M}$, we selected two proprietary and two open-weights models. GPT and Claude were the dominant models observed in our GitHub dataset. Hence we chose \texttt{gpt-5.6-sol} and \texttt{sonnet-5}. For the open-weights models, we selected \texttt{deepseek-v4-pro} and \texttt{qwen-3.8-max}, as DeepSeek and Qwen models have been commonly used in prior studies on software accessibility~\cite{aljedaani2026multi, suh2026empirical}. The selected variants represent default (for proprietary models) and frontier (for the open-weights models) versions of their respective model families at the time of writing this paper. 

For agents $\mathcal{A}$, we selected Codex, Claude Code and OpenCode. The first two represent the most popular proprietary coding agents, while OpenCode serves as an open source alternative, used in agentic software engineering research~\cite{chen2026failedtrajectoriesreliablellm, shu2026resolveratehidestrajectory}.

Since these are all reasoning models, we were able to configure their reasoning effort levels $\mathcal{E}$ as well. The levels include: default, high, medium and low. This determined how much reasoning effort would be applied by the models to work on the given prompt.

\subsubsection{Skills}
\label{subsubsec:skills}

Skills are a form of agentic intervention that guide agents in conducting a specified task~\cite{anthropic2026agentskills}. A Skill lives in an agent's workspace as a markdown file (\texttt{Skill.md}), and is probabilistically invoked by the agent if a task matches the description and condition written in that file. The Skill can include optional resources and provides guidance, context and tool support.

For our experiments, we looked for Skills designed to support web accessibility-related tasks, such as WCAG auditing, ARIA and semantic-HTML remediation, screen-reader and keyboard support, automated accessibility testing, and more. Our Skill retrieval process consisted of the following stages.
\begin{enumerate}
    \item Manual search: We searched in GitHub and Skill marketplaces for relevant Skills using ``accessibility'' as the keyword. We listed 15 Skills as relevant ground truth set.
    
    \item Keyword-scoped discovery on GitHub: Our manual search indicated a more vast collection of Skills hosted and distributed on GitHub. Therefore, we expanded our search in GitHub, retrieving repositories that contained a ``SKILL.md'' file with the mention of ``WCAG''. We found 74,577 Skills in 22,286 repositories.

    \item Lexical filter: We scored each Skill on the occurrence of 21 accessibility-related terms, including wcag, accessibility, accessible, a11y \footnote{The term `a11y' is a numeronym, shortened version of the word `accessibility'}, aria, aria-label, screen reader, nvda, jaws, voiceover, color contrast, alt text, keyboard navigation, focus management, assistive technology, section 508, axe-core, pa11y, lighthouse accessibility, inclusive design, POUR principles. We filtered Skills with $\geq4$ terms, shortening our dataset to 39,210 Skills in 14,477 repositories.
    
    \item Deduplication: Skills can be shared and adopted across many repositories. Hence, the name, description and content in the \textit{Skill.md} files were compared to detect duplicates and merge them as one unique Skill. The resulting number of unique Skills was 15,344.

    \item Relevance filter: Next, we employed an LLM (Claude Sonnet-5) to classify the relevance of the Skills, so that we can filter Skills whose core purpose is web accessibility. Exclusion criteria include: (i) general UI/UX, frontend, testing, or design-system skills that list accessibility as one concern among several; (ii) native-only accessibility (iOS, Android, SwiftUI, WPF, desktop), which is outside our web scope; (iii) PDF, Office, and document accessibility; (iv) audits that give equal weight to accessibility and to performance, security, or visual polish; and (v) uses of ``accessible'' as a general quality adjective (e.g. ``accessible pricing''). We used our ground truth data from step 1 to verify the scope of the exclusion criteria. This reduced the dataset to 3,658 Skills (4,138 repositories).

    \item Ranking: As the final step, we shortlisted the top 50 Skills based on popularity. Popularity was measured by how many unique repositories contained that Skill. These 50 represent the most commonly used Skills for improving accessibility.
\end{enumerate}

The final Skills were then manually annotated for stage, coverage, trigger and resources. This content analysis~\cite{krippendorff2018content} was conducted with independent iterative open coding by two authors on 10 Skills (20\% of the dataset). Disagreements were discussed and remediated. One author coded the remaining Skills, with the second reviewing the codes afterwards. An overview of the coding dimensions and labels is provided in Table \ref{tab:method-artifacts-skills}. 

\begin{table*}[t]
    \centering
    \small
    \caption{Overview of skill annotation dimensions and corresponding codes.}
    \label{tab:method-artifacts-skills}

    \begin{tabularx}{\textwidth}{
        >{\raggedright\arraybackslash}l
        >{\raggedright\arraybackslash}X
        >{\raggedright\arraybackslash}p{0.4\textwidth}
    }
        \toprule
        \textbf{Dimension} &
        \textbf{Description} &
        \textbf{Codes} \\
        \midrule

        Stage &
        Captures the primary intent of the skill, i.e. at which stage of the software development cycle is the skill's intervention intended for. &
        Prevent/guide, Audit/detect, Fix/remediate, and Report \\

        Trigger &
        Describes the condition or user action that initiates the use of the skill. &
        Prompt, Command, and Implicit \\

        Scope &
        Indicates the extent of accessibility specification in the skill. &
        Generic, Specific issues, and Contrast \\

        Resources &
        Indicates the types of resources included within and without the skill. &
        Markdown files, Code snippets, Linked tools, Linked documentation, and Scripts \\

        \bottomrule
    \end{tabularx}
\end{table*}

\subsection{Measures}
\label{subsec:measure}

\subsubsection{Accessibility Evaluation}
\label{subsubsec:a11y}
To calculate the primary goal of this study --- accessibility performance --- we conducted automated accessibility audits with three tools and converged their individual findings to generate an accessibility score for each website instance.

\begin{figure}[h]
    \centering
    \includegraphics[width=.8\columnwidth]{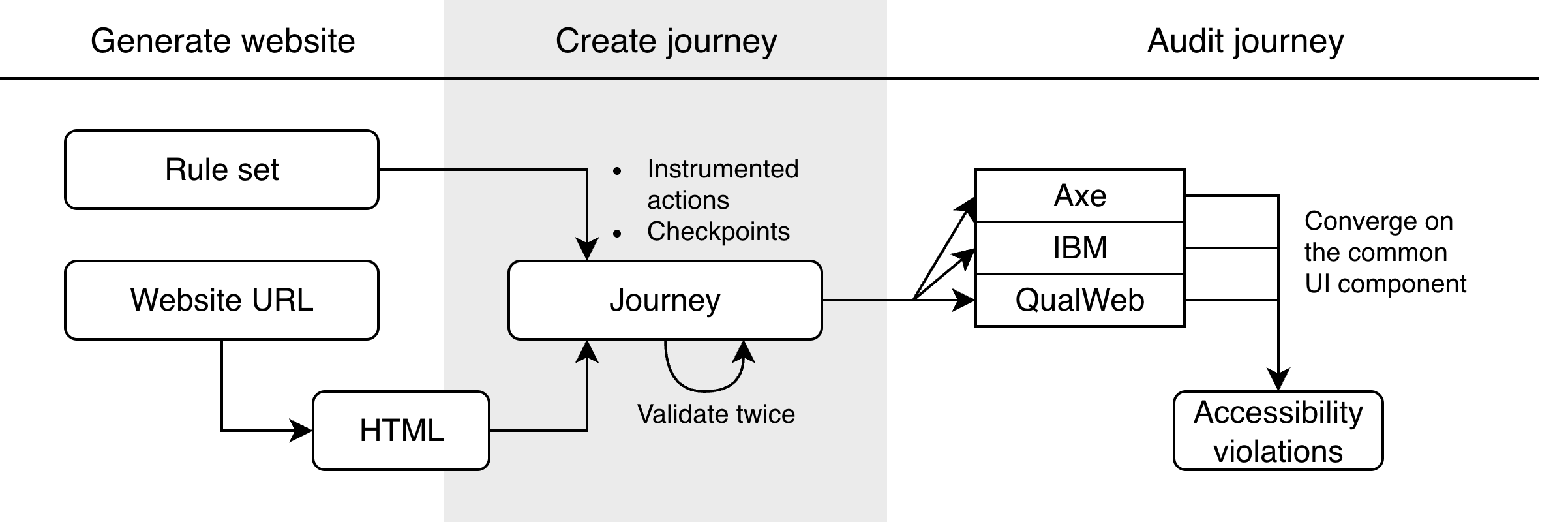}
    \caption{Overview of the process of accessibility evaluation}
    \Description{A flowchart with three stages: "Generate Website", "create journey" and "Audit journey". Rule sets and HTMLs from Website URLs are retrieved from the first stage and sent to create Journeys in the second stage. Journeys include Instrumented actions and Checkpoints, and are validated twice. Journeys are then sent to three tools: Axe, IBM and QualWeb in the last stage. From them, reports converge on the common UI component to end up as a list of Accessibility violations.}
    \label{fig:a11y_audit}
\end{figure}

The process of accessibility evaluation, as shown in Fig~\ref{fig:a11y_audit}, began when a website instance was generated by an agent and uploaded to GitHub Pages for deployment. First, a \texttt{journey} was created for that instance based on its project rule set $R_p$. A journey is a sequence of mechanical actions on a website interface (e.g., finding a specific component, writing in a text field, pressing a button, etc.) with the goal of accomplishing a task. These journeys were created using Playwright~\cite{microsoft2026playwright}, exposed to an agent (Claude Sonnet-5 in our experiment) through an MCP~\cite{mcp2026intro}. The agent would read the rule set and perform instrumentation on the provided website, writing the corresponding action sequence.

Journeys would also include \texttt{checkpoints}, where a snapshot of the website in its current state was taken. This snapshot was then used for auditing.

After journeys were created, three open-source automated web accessibility evaluation tools were used to audit them. These tools, often used in literature~\cite{mowar2025codea11y, aljedaani2024does, suh2026empirical}, are axe-core (\textit{Axe})~\cite{axecore}, IBM Equal Access Accessibility Checker (\textit{IBM})~\cite{ibmequalaccess}, and \textit{QualWeb}~\cite{qualweb}. These rules-based testers evaluate the HTML snapshots for violations of WCAG Success Criteria and ACT rules. 

Each tool generated a list of UI components with corresponding violations. These separate reports were then converged based on the underlying component. From the single converged report, we calculate the accessibility score $S_{a11y}$,

\begin{equation}
\label{eq:a11y}
S_{a11y} = 100\left(1 - \frac{VP + CB}{2}\right)
\end{equation}

where $VP$ represents \textit{Violation Prevalence} and $CB$ represents \textit{Criterion Breadth}. 

Violation prevalence captures the rate of accessibility violation instances in a website, defined as: 
\begin{equation}
\label{eq:vp}
VP = \frac{N_{\mathrm{vUI}}}{N_{\mathrm{tUI}}}
\end{equation} 
where $N_{\mathrm{vUI}}$ is the number of UI elements associated with detected accessibility violations and $N_{\mathrm{tUI}}$ is the number of total UI elements in the website. 

In some cases, the prevalence of a single type of issue can bloat VP. To counteract this, we included criterion breadth in our scoring, which checks how many distinct WCAG Success Criteria (SCs) were violated. It is defined as: 
\begin{equation} 
\label{eq:cb}
CB = \frac{N_{\mathrm{vSC}}}{N_{\mathrm{tSC}}}
\end{equation} 
where $N_{\mathrm{vSC}}$ is the number of unique WCAG SCs violated, and $N_{\mathrm{tSC}}$ is the total number of SCs covered by the automated tools.

With a range $[0, 100]$, higher $S_{a11y}$ values indicate fewer violated UI components and less SCs impacted. Hence, for our experiments, the higher the $S_{a11y}$, the more accessible the website.

On top of automated testing, we conducted user tests with blind individuals. The goal of the tests was to complement our automated results, map static failures to experienced problems, and strengthen claims on accessibility. We recruited three blind testers through program-l, a mailing list for blind and visually impaired professionals~\cite{ProgramL}. Each participant was given the same set of 6 websites, two from each project. Each pair included one instance generated with no Skill, and the same configuration generated with a Skill (more details in Section~\ref{sec:rq2}). The exact configurations for the instances were dictated by their relative $S_{a11y}$. We selected a random project from the highest scoring quartile for each project.

The testers were instructed to complete the same set of tasks, influenced by the project's rule set. For instance, the first task for $P_2$-Planner was to create a new recipe. They were encouraged, but not required, to think aloud~\cite{VanSomeren1994Think}. Assistance on challenging tasks was not provided immediately, so that they could understand and report accessibility issues more accurately. After task completion, the tester was asked about their experience, elaborating on any issue faced. These confirmed feedback were noted down as issue instances. These test sessions were approved by the IRB of the author's institution, and participants were compensated with 60 USD gift cards for 1.5 hours long sessions.

\subsubsection{Functional Correctness}
\label{subsubsec:func}
As a secondary measurement of agent performance on creating websites, we also evaluated functional correctness. Prior work~\cite{zhong2025swe, zhao2025vibe} have shown how additional constraints on requirements such as style and security can degrade functional correctness. 

Similar to accessibility evaluation, functional correctness was measured on the generated journeys. Assertions were written based on the rule set $R_p$. The score is calculated as,

\begin{equation}
S_{func} = 100\left(1 - \frac{N_{fail}}{N_{total}}\right)
\end{equation}

where $N_{\mathrm{fail}}$ is the number of test cases failed and $N_{\mathrm{total}}$ is the total number of test cases. Higher $S_{func}$ corresponds to better functional correctness.

To verify the accuracy of automated functionality testing, a subset of website instances was manually tested. The testing was conducted independently by two 4th year Software Engineering undergraduate students who are not authors of this paper. Three instances per project were selected, with each subset including one Codex, one Claude, and one OpenCode-generated implementation. These also varied in underlying model and reasoning-effort configuration. Testers were provided with the URLs of the nine websites, and test cases from the corresponding project's rule set.

For each test case, testers recorded a categorical outcome of Pass, Fail or Unsure, and optionally provided supporting evidence or notes. Inter-rater agreement was calculated over the shared test cases. The two manual testers agreed on 118 of 129 test cases, corresponding to 91.5\% exact agreement. Eleven disagreements were reviewed by a third reviewer, who produced a complete consensus set without Unsure outcomes. Comparing the consensus results with the automated results, we observed an agreement of 97.7\% (126/129), indicating a satisfactory automation process.

\subsubsection{Trajectories}
\label{subsubsec:trajectory}
In recent times, analyzing agent success has shifted from evaluating only the outcome into analyzing their trajectories~\cite{bouzenia2025understanding, shu2026resolveratehidestrajectory, liu2026codemind, chen2026failedtrajectoriesreliablellm}. A trajectory is the sequence of agent actions. Actions include navigating folders, reading files, planning, writing and editing code, installing libraries, deploying and testing websites, running verification tools, and more. Optionally, actions can be associated with reasoning summaries, indicating what the agent reasoned about before taking an action.
For our experiments, we analyzed three components of the trajectory:
\begin{itemize}
    \item Trajectory phases: we divided the trajectory into multiple phases based on action intent. All agents followed a planning - implementation - verification phase sequence.
    
    \item Accessibility-mention in reasoning: we extracted reasoning summaries and inspected whether accessibility or accessibility-related terminologies such as ARIA, skip link, color contrast, etc. were mentioned. The goal was to observe how much agents reasoned about accessibility.

    \item Accessibility verification actions: we checked the actions in verification phases to see whether it is relevant to accessibility. Running an accessibility evaluation tool and executing a color contrast script are examples of such actions. 
\end{itemize}

Each component was automatically detected with deterministic heuristics that was verified and refined through independent manual annotation. After multiple rounds of annotation, heuristics were finalized once agreement with manual codes reached a $\kappa \geq 0.8$. The manual coding workbooks are provided in the replication package for reference~\cite{replication_package}.

\section{RQ1 - Accessibility of LLM Coding Agents}
\label{sec:rq1}

\subsection{Experimental Setup}
\label{subsec:rq1-exp}

To answer RQ1, we generated a total of 90 website instances $W_{p,c}^{(r)}$, where $p$ denotes the project $P_p$, $c$ the configuration of coding agent $\mathcal{A}$, model $\mathcal{M}$ and reasoning effort level $\mathcal{E}$, and $r$ the replication version.
To cover different combinations of models and agents, and check the impact of effort levels, we formulated 22 different configurations across 3 pools so that 
$
\mathcal{C}_{\mathrm{RQ1}}
=
\mathcal{C}_{base}
\cup
\mathcal{C}_{agent}
\cup
\mathcal{C}_{effort},
$
where \(\mathcal{C}_{base}\), \(\mathcal{C}_{agent}\), and \(\mathcal{C}_{effort}\) denote the following three configuration pools. 

\begin{itemize}
\item Base pool (\(\mathcal{C}_{base}\), 4 configurations):
Codex--GPT--default,\footnote{Going forward, for readability, models are represented by shortened names: \texttt{gpt-5.6-sol}=GPT, \texttt{sonnet-5}=Sonnet, \texttt{deepseek-v4-pro}=DeepSeek, and \texttt{qwen-3.8-max}=Qwen.}
Claude--Sonnet--default,
OpenCode--DeepSeek--default, and
OpenCode--Qwen--default.

\item Agent-variable pool (\(\mathcal{C}_{agent}\), 12 configurations):
each of the three agents was paired with each of the four models at high reasoning effort,
\[
\mathcal{C}_{agent}
=
\mathcal{A}
\times
\mathcal{M}
\times
\{\mathrm{high}\}.
\]

\item Effort-variable pool (\(\mathcal{C}_{effort}\), 9 configurations):
OpenCode was paired with GPT, Sonnet, and Qwen at high, medium, and low reasoning effort,
\[
\mathcal{C}_{effort}
=
\{\mathrm{OpenCode}\}
\times
\{\mathrm{GPT},\mathrm{Sonnet},\mathrm{Qwen}\}
\times
\{\mathrm{high},\mathrm{medium},\mathrm{low}\}.
\]
DeepSeek was excluded from this pool because it did not expose a medium reasoning-effort level through OpenCode.
\end{itemize}


Since LLMs are non-deterministic, we constructed replications of each configuration in the base pool $\mathcal{C}_{base}$ twice. We denote the replicate index by $r \in \{1,2,3\}$, where \(r=1\) denotes the original generation and \(r=2,3\) denote the two additional replicates.

\subsection{Results}
\label{subsec:rq1_results}

We evaluated the resulting 90 website instances for accessibility with automated evaluators (Section~\ref{subsubsec:a11y}), and calculated their Accessibility Scores. We also checked how functionally correct they were, based on automated functionality testing (Section~\ref{subsubsec:func}). Of the 90 runs, 80 scored a perfect 100 with the average $S_{func}$ of 97.83. However, None of the 90 runs, across 3 projects, 22 different configurations and 12 replicates, was fully accessible. Specifically, with an average accessibility score of 88.33, coding agents generated websites with an average 80.43 violation instances and 3.61 distinct WCAG SC violations. The project-wise breakdown is shown in Table~\ref{tab:rq1-result-01}. Results for all website instances, along with their URLs, are provided in the replication package~\cite{replication_package}.

\begin{table*}[t]
  \centering
  \small
  \caption{Project-wise breakdown of accessibility scores (out of 100), along with the number of UI components with confirmed violation ($N_{vUI}$), and distinct WCAG 2.2 Success Criteria (SC) violations ($N_{vSC}$).}
  \label{tab:rq1-result-01}
  \begin{tabular}{l*{9}{r}}
    \toprule
    \multirow{2}{*}{Project}
      & \multicolumn{3}{c}{Accessibility Score}
      & \multicolumn{3}{c}{$N_{vUI}$}
      & \multicolumn{3}{c}{$N_{vSC}$} \\
    \cmidrule(lr){2-4}\cmidrule(lr){5-7}\cmidrule(lr){8-10}
      & Avg. & Max. & Min. & Avg. & Max. & Min. & Avg. & Max. & Min. \\
    \midrule
    $P_1$-Catalog
      & 88.95 & 97.24 & 73.87
      & 61.37 & 178 & 2
      & 4.43 & 7 & 2 \\
    $P_2$-Planner
      & 86.17 & 95.79 & 74.84
      & 144.53 & 425 & 15
      & 3.67 & 8 & 2 \\
    $P_3$-Viz
      & 89.87 & 96.13 & 80.90
      & 35.40 & 80 & 4
      & 2.73 & 5 & 1 \\
    \midrule
    Overall
      & 88.33 & 97.24 & 73.87
      & 80.43 & 425 & 2
      & 3.61 & 8 & 1 \\
    \bottomrule
  \end{tabular}
\end{table*}

Inaccessibility was also corroborated by our blind user testers. In Table~\ref{tab:rq1-results-usertest}, we show the number of issues reported by each tester.

\begin{table}[t]
  \centering
  \small
  \caption{Number of accessibility issues identified by blind testers T1, T2 and T3, along with the number of violated UI components ($N_{vUI}$) and number of distinct WCAG 2.2 Success Criteria violations ($N_{vSC}$) detected by our automated process.}
  \label{tab:rq1-results-usertest}
  \begin{tabular}{l>{\raggedright\arraybackslash}l*{5}{r}}
    \toprule
    Project & Configuration & T1 & T2 & T3 & $N_{vUI}$ & $N_{vSC}$ \\
    \midrule
    $P_1$-Catalog & Codex / DeepSeek / high & 4 & 2 & 3 & 6 & 4 \\
    $P_2$-Planner   & Claude / Sonnet / high   & 10 & 7 & 5 & 15 & 2 \\
    $P_3$-Viz & OpenCode / GPT / high     & 2 & 2 & 3 & 25 & 2 \\
    \bottomrule
  \end{tabular}
\end{table}

\subsubsection{Accessibility Issues}

We investigated the exact accessibility issues reported, to check the agents' weaknesses. We observed that 13 SCs appear at least once in our runs. In Table~\ref{tab:rq1-results-sc}, we show the number of runs with SC violations along with the average number of UI components with that violation. The most pervasive issue was 1.4.3 Contrast (Minimum), which was violated in all instances in all projects, with the highest average instances of 62.71.

\begin{table}[t]
  \centering
  \small
  \caption{Distribution of violated WCAG success criteria across projects, sorted by total number of occurences.}
  \label{tab:rq1-results-sc}
  \begin{tabular}{ll*{5}{c}}
    \toprule
    \multicolumn{2}{c}{Success Criterion} & $P_1$-Catalog & $P_2$-Planner & $P_3$-Viz & Total & Avg UI Violations \\
    \midrule
    1.4.3 & Contrast (Minimum) & 30/30 & 30/30 & 30/30 & 90 & 62.71 \\
    4.1.2 & Name, Role, Value & 16/30 & 30/30 & 12/30 & 58 & 14.86 \\
    1.3.1 & Info and Relationships & 15/30 & 17/30 & 5/30 & 37 & 5.43 \\
    2.5.3 & Label in Name & 17/30 & 10/30 & 9/30 & 36 & 8.69 \\
    3.1.1 & Language of Page & 29/30 & 0/30 & 0/30 & 29 & 1.00 \\
    2.1.1 & Keyboard & 20/30 & 2/30 & 0/30 & 22 & 1.32 \\
    1.1.1 & Non-text Content & 2/30 & 2/30 & 17/30 & 21 & 7.00 \\
    2.5.8 & Target Size (Minimum) & 0/30 & 13/30 & 0/30 & 13 & 10.00 \\
    2.4.1 & Bypass Blocks & 1/30 & 1/30 & 9/30 & 11 & 2.18 \\
    2.4.7 & Focus Visible & 1/30 & 2/30 & 0/30 & 3 & 3.33 \\
    3.2.1 & On Focus & 1/30 & 2/30 & 0/30 & 3 & 3.33 \\
    2.4.4 & Link Purpose (In Context) & 1/30 & 0/30 & 0/30 & 1 & 12.00 \\
    3.3.2 & Labels or Instructions & 0/30 & 1/30 & 0/30 & 1 & 1.00 \\
    \bottomrule
  \end{tabular}
\end{table}

We report on the recurring issues on each project, as they represent similar implementations.

$P_1$-Catalog websites created a catalog of course cards, and a split-view course details page with the left side for syllabus and right side for material-viewing. Contrast issues appeared repeatedly on the course and syllabus table: week/date/content headers, date cells, material buttons, and secondary labels such as “PDF Reading”. Visible-label and accessible-name mismatches appeared in 16/30 instances. A clear example was a brand button whose visible text was “EpochHistory, in context,” but whose accessible name was set to “Go to course catalog.” Separately, keyboard-inaccessible scrollable regions appeared in 18/30 configurations. 
  
$P_2$-Planner websites implemented a gallery of meals, a detailed form for creating new recipes, a calendar view to display and operate a meal planner, and a shopping list of ingredients containing items necessary for meals included in the plan. The most common issue, color contrast, occurred in the top navigation tabs as well as in recipe-cards, similar to the previous project. The contrast issue was propagated in the planner and the shopping list, suggesting a shared color-token problem affecting secondary text and button styles, rather than an isolated defect in one component.

Form semantics were the second major pattern. Inputs and selects lacked reliable accessible names or associated labels in 27/30 configurations. Small or insufficiently spaced controls appeared in 13/30 configurations, especially repeated add, remove, quantity, and meal-slot buttons. Unnamed buttons appeared in 8/30 configurations, unlabeled complementary regions in 11/30, and malformed grid or menu ARIA structures in 6/30. Overall, the project’s recurring weaknesses were the shared contrast system and the semantics and sizing of its dense, repeated form controls.
  
Lastly, $P_3$-Viz websites generated an interactive list of travel cards, showing the itinerary of a trip, along with an interactive pie chart of the trip expenses with corresponding bar charts showing category-specific breakdown. The prevalent contrast problems appeared in itinerary tables, trip descriptions, and chart-related text. Chart and SVG accessibility defects appeared in 25/30 configurations, including missing accessible names, inappropriate image or application roles, invalid descendants, and interactive content nested inside graphical elements. Visible-label and accessible-name mismatches affected navigation and day-selection controls in 9/30 configurations. The dominant project-specific problem was therefore that visualizations were visually rich but were not consistently represented to assistive technology.

\subsubsection{Configurations}

Next, we explored whether the choice of configuration --- coding agent, model and effort level --- influences accessibility performance. Coding agent and model associations were evaluated on the Agent-variable pool (number of instances$=36$) while effort used the Effort-variable pool ($n=27$). For statistical analysis, we fitted regression-based factorial ANOVA models. 

\begin{figure*}[t]
  \centering
  \begin{subfigure}[t]{0.32\textwidth}
    \centering
    \includegraphics[width=\linewidth]{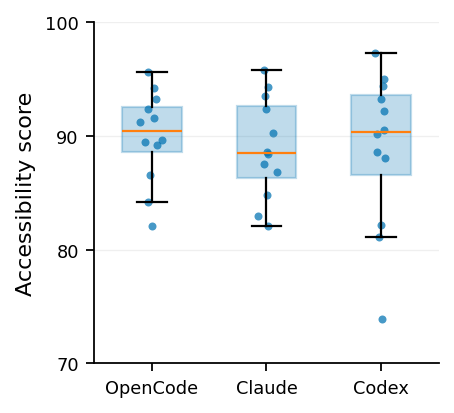}
    \caption{$S_{a11y}$ distribution by agent.}
    \Description{A box plot with Accessibility Score on the y-axis ranging from 70 to 100, and three agents in the x-axis: OpenCode, Claude and Codex.}
    \label{fig:rq1-results-agent}
  \end{subfigure}\hfill
  \begin{subfigure}[t]{0.32\textwidth}
    \centering
    \includegraphics[width=\linewidth]{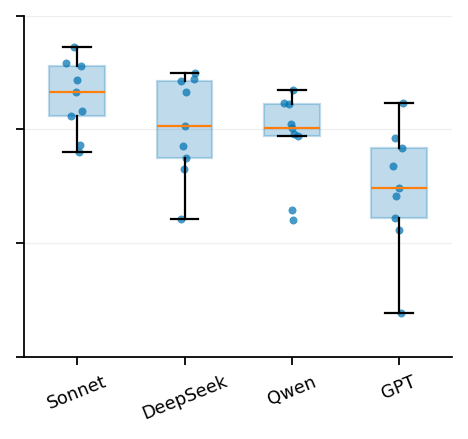}
    \caption{$S_{a11y}$ distribution by model.}
    \Description{A box plot with Accessibility Score on the y-axis ranging from 70 to 100, and four models in the x-axis: Sonnet, DeepSeek, Qwen and GPT.}
    \label{fig:rq1-results-model}
  \end{subfigure}\hfill
  \begin{subfigure}[t]{0.32\textwidth}
    \centering
    \includegraphics[width=\linewidth]{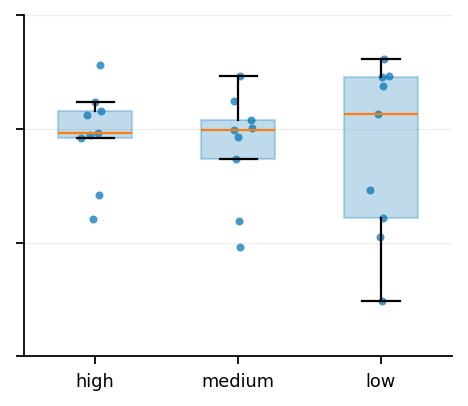}
    \caption{$S_{a11y}$ distribution by effort.}
    \Description{A box plot with Accessibility Score on the y-axis ranging from 70 to 100, and three effort levels in the x-axis: high, medium, low.}
    \label{fig:rq1-results-effort}
  \end{subfigure}
  \caption{Accessibility Scores ($S_{a11y}$) by agent, model, and reasoning-effort level. Points represent individual project runs; boxes summarize the score distributions.}
    \Description{A collection of three box plots, each sharing Accessibility Score as the y-axis ranging from 70 to 100, and showing coding agents, models and reasoning-effort levels in the in the x-axis respectively.}
  \label{fig:rq1-results-config}
\end{figure*}

Distributions of accessibility scores by agent, model and effort are provided in Fig.~\ref{fig:rq1-results-config}. For coding agents, we observe that mean accessibility scores were similar for Claude ($Mean=88.94$), Codex ($M=88.88$), and OpenCode ($M=89.94$). No statistically significant association between agent choice and accessibility score was detected ($p=.823$).

For models, Sonnet had the highest mean accessibility score ($M=92.85$). The open-weights models, DeepSeek ($M=90.20$) and Qwen ($M=89.19$), were not far behind. But GPT ($M=84.77$) scored the lowest. The statistical analysis showed a significant association with model choice ($p=.013$) with a large effect size (partial $R^2=.346$).

The level of reasoning effort put high effort with best mean performance ($M=89.47$), compared with medium ($M=88.44$) and low effort ($M=88.06$). A low reasoning effort shows wider distribution of performance. However, no statistically significant association between reasoning-effort level and accessibility score was detected ($p=.823$).

\subsubsection{Trajectory analysis}

We wanted to further understand agentic runs through their trajectories and derive any relationship with accessibility outcomes. From the extracted and annotated trajectories (as described in Section~\ref{subsubsec:trajectory}), we looked into action counts, phases, and reasoning. Associations were evaluated using two-sided Spearman rank correlations.

\begin{figure*}[t]
  \centering
  \includegraphics[width=0.72\textwidth]{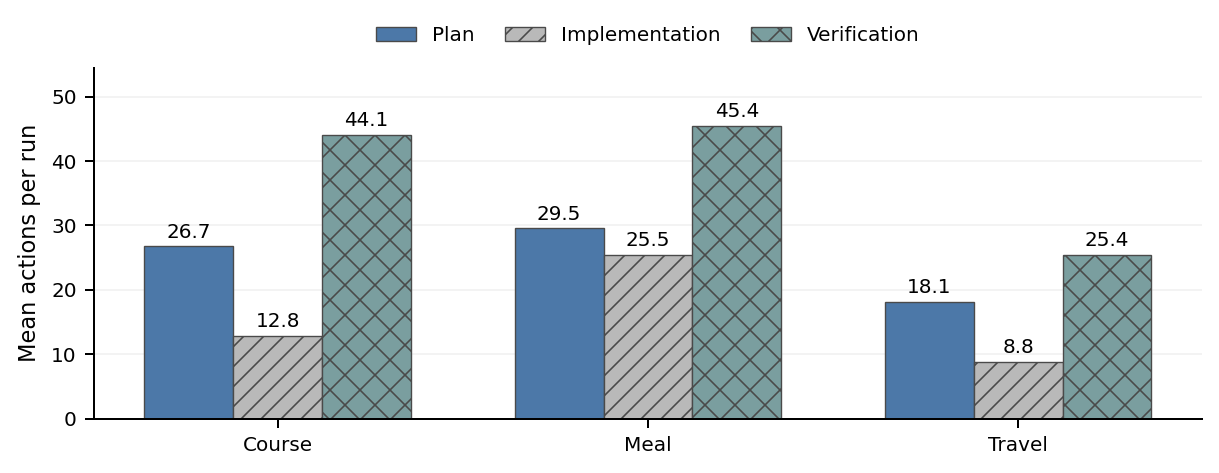}
  \caption{Mean phase-level semantic action counts separated by project (30 runs per project).}
    \Description{Clustered bar charts, each representing a project. Each cluster contains three bars representing the phases Plan, Implementation and Verification. The y axis shows mean actions per run. In all projects, verification has the highest numbers.}
  \label{fig:rq3-baseline-phase-actions}
\end{figure*}

RQ1 runs ($n=90$) contained an average of 78.79 actions ($SD=51.05$): 24.80 during planning ($SD=17.18$), 15.69 during implementation ($SD=14.82$), and 38.30 during verification ($SD=30.89$). Thus, majority of observed actions occurred in the final verification phase. Figure~\ref{fig:rq3-baseline-phase-actions} separates these phase-level means by project. 

Based on our statistical analysis, verification-phase action count had a moderate association with a higher accessibility score ($\rho=.307$, $p=.003$) and lower violation prevalence ($\rho=-.331$, $p=.001$). This is an association with the amount of verification activity, not evidence that additional tool calls themselves caused better accessibility, since our annotation revealed no accessibility-specific verification actions in any run.

Readable reasoning summaries were exposed for 81 of the 90 runs.  Accessibility-related reasoning occurred in 39 of these 81 runs (48.1\%), comprising 86 summaries in total and a mean of 2.21 summaries among runs in which it occurred. Reasoning primarily concerned accessible names and ARIA (SC~4.1.2), semantic structure (SC~1.3.1), text alternatives (SC~1.1.1), keyboard access (SC~2.1.1), and visible focus (SC~2.4.7). Reasoning count was not significantly associated with accessibility score ($\rho=.212$, $p=.057$) or violation prevalence ($\rho=-.202$, $p=.070$). Runs with any accessibility reasoning averaged 89.73 in accessibility score and 11.67\% violation prevalence, compared with the overall averages of 88.61 and 13.83\%, respectively.

\subsubsection{Run-to-run variability}

We analyzed the replicates ($n=12 \times 3$) on accessibility score, functional correctness and trajectory size to examine run-to-run variability of exact configurations. Variability was defined as the range between the highest and lowest replicate value.

Across configurations, accessibility score had a mean three-run range of 6.60 points (median 4.61). Functional correctness had a mean range of 7.43 points (median 2.78); six of the 12 configurations were unchanged across replicates. Trajectory measures were more variable: total token use differed by an average of 1.80 million tokens between the lowest- and highest-token replicate (median 0.92 million), and recorded event count had a mean range of 24.50 events (median 23.50).

\section{RQ2 - Accessibility Improvement with Interventions}
\label{sec:rq2}

\subsection{Experimental Setup}

To answer RQ2, we selected two Skills from our set of 50 mined Skills as described in Section~\ref{subsubsec:skills}.

\begin{itemize}
    \item accessibility-audit ($SK_{gen}$): From our content labeling of the 50 Skills, we found that the majority of Skills were created for the \textit{audit/detect} stage, included \textit{code snippets}, \textit{linked tools} and \textit{linked documentation} as resources, and encompassed a \textit{generic} scope, covering multiple WCAG 2.2 SCs. We selected this Skill to represent these three characteristics, the first of which is also supported by prior work~\cite{aljedaani2026multi, suh2026empirical}, where iterative testing was shown to improve accessibility.

    \item contrast-checker ($SK_{con}$): Additionally, we wanted to see whether \textit{specific issues}-scoped Skills with purpose-driven \textit{scripts} would perform better than general-purpose ones on that specific issue. Given that color contrast was the most prevalent issue observed in RQ1, we selected this Skill with 3 dedicated contrast checking scripts, as our second intervention.
\end{itemize}

We renamed the Skills and edited their descriptions for better discoverability. Our content analysis showed that almost all Skills relied on explicit \textit{prompt} or \textit{command}-based trigger. However, the purpose of our study is to evaluate implicit interventions. Therefore, we modified the text so that the Skill can be autonomously invoked consistently across different instances, during the verification stage after building the website.

We designed RQ2 to compare websites generated with Skills against the corresponding no-Skill website, generated under the same project and configuration. We define three skill conditions, $\mathcal{V}=\{v_1,v_2,v_3\}$, where $v_1=\text{no skill}$, $v_2=SK_{gen}$, and $v_3=SK_{con}$.

We selected a subset of 8 configurations per project based on accessibility scores, 2 per quartile. In total, we generated additional 24 \(v_2\) and 24  \(v_3\) website instances for RQ2.


\subsection{Results}

\subsubsection{General accessibility Skill intervention}

First, we demonstrate the comparative results between \(v_1\) and \(v_2\) accessibility scores. As shown in Table~\ref{tab:rq2-results-01-v2}, across the 24 matched \(v_1\)-\(v_2\) pairs, mean accessibility score increased from 88.06 to 92.13. The mean improvement of 4.07 was statistically significant using a paired $t$-test, $t(23)=3.20$, $p=.004$, with a moderate paired effect size
($d_z=.65$). Improvements are similarly observed from our user studies. Table~\ref{tab:rq2-results-usertest} lists the number of issues detected by blind testers. Out of 9 test instances ($project \times tester$), 5 showed improvement, 3 stayed the same, and 1 caused an increase in inaccessibility. However, it is to be noted that despite the improvements, no instance reached a perfect score.

\begin{table*}[t]
  \centering
  \small
  \caption{Matched accessibility-score changes after inclusion of the generic skill.}
  \label{tab:rq2-results-01-v2}
  \begin{tabular}{lrrrrc}
    \toprule
    Project & $n$ & \(v1\), mean (SD) & \(v2\), mean (SD) & Mean change & Improved / worsened \\
    \midrule
    $P_1$-Catalog  & 8  & 91.09 (4.86) & 90.44 (6.94) & $-0.65$ & 4 / 4 \\
    $P_2$-Planner    & 8  & 84.22 (7.93) & 91.40 (5.87) & $+7.18$ & 7 / 1 \\
    $P_3$-Viz  & 8  & 88.87 (4.34) & 94.54 (2.57) & $+5.67$ & 8 / 0 \\
    \midrule
    Overall & 24 & 88.06 (6.37) & 92.13 (5.51) & $+4.07$ & 19 / 5 \\
    \bottomrule
  \end{tabular}
\end{table*}

\begin{figure*}[t]
  \centering
  \includegraphics[width=0.86\textwidth]{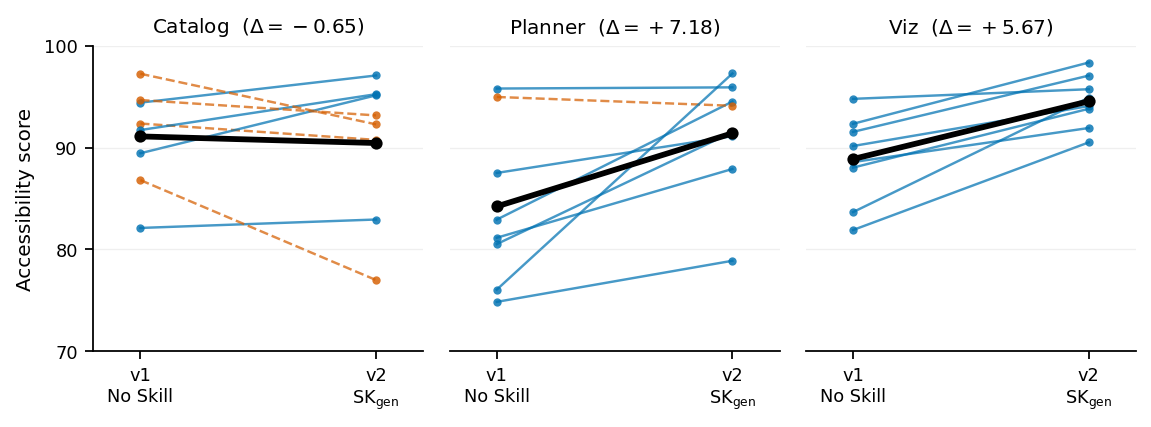}
  \caption{Matched accessibility-score changes following inclusion of the general Skill $SK_{gen}$. Solid blue lines indicate improving runs and dashed orange lines indicate worsening runs; thick black lines connect project means. All panels use the same truncated 70--100 y-axis.}
  \Description{Three paired-slope graphs, each representing a project. The y-axis is accessibility score, and x axis has two values: v1 No Skill and v2 General Skill. The lines demonstrate change in score between pairs of the same configuration between v1 and v2. There are 19 improving lines and 5 regressing. The mean line is downward for Catalog but upward for Planner and Viz projects.}
  \label{fig:rq2-results-01-v2}
\end{figure*}

Furthermore, improvement was not uniform, with 5 out of 24 pairs with a decreasing accessibility score. Fig.~\ref{fig:rq2-results-01-v2} shows all the paired scores in the three projects. For $P_1$-Planner, \(v_2\) instances contained similar contrast failures and naming mismatch as \(v_1\). The other two projects saw substantial improvements, by 7.18 and 5.67 points respectively.

\begin{table}[t]
  \centering
  \small
  \caption{Number of accessibility issues identified by blind testers.}
  \label{tab:rq2-results-usertest}
  \begin{tabular}{l*{6}{r}}
    \toprule
    \multirow{2}{*}{Project}
      & \multicolumn{2}{c}{Tester T1}
      & \multicolumn{2}{c}{Tester T2}
      & \multicolumn{2}{c}{Tester T3} \\
    \cmidrule(lr){2-3}\cmidrule(lr){4-5}\cmidrule(lr){6-7}
      & \(v1\) & \(v2\) & \(v1\) & \(v2\) & \(v1\) & \(v2\) \\
    \midrule
    Catalog & 4 & 2 & 2 & 1 & 3 & 3 \\
    Planner   & 10 & 9 & 7 & 5 & 5 & 6 \\
    Viz & 2 & 2 & 2 & 1 & 3 & 3 \\
    \bottomrule
  \end{tabular}
\end{table}

\subsubsection{Contrast-specific Skill intervention ($SK_{con}$)}

To evaluate $SK_{con}$'s effectiveness in mitigating contrast issues, we calculated SC 1.4.3 Violation Prevalence ($VP_{SC1.4.3}$) following Equation~\ref{eq:vp} such that only SC 1.4.3 violations were considered. We report the percentage value for better readability.

\begin{figure*}[t]
  \centering
  \includegraphics[width=0.88\textwidth]{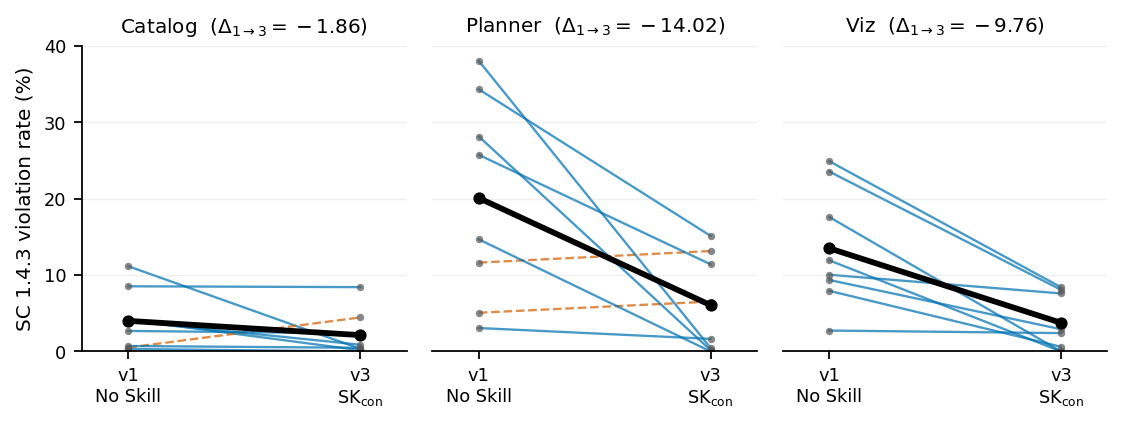}
  \caption{Matched SC~1.4.3 violation-prevalence comparison following inclusion of the SC 1.4.3-specific Skill $SK_{con}$. Solid blue lines indicate an improving lower prevalence and dashed orange lines the opposite. Thick black lines connect project means.}
  \Description{Three paired-slope graphs, each representing a project. The y-axis is accessibility score, and x axis has two values: v1 No Skill and v3 Contrast Skill. The lines demonstrate change in score between pairs of the same configuration between v1 and v3. There are 21 improving lines and 3 regressing. The mean lines are upward for all three projects.}
  \label{fig:contrast-skill-v1-v3}
\end{figure*}

Results indicate a significantly positive outcome. Fig.~\ref{fig:contrast-skill-v1-v3} shows $VP_{SC1.4.3}$ pairs between \(v1\) and \(v3\), where 21/24 have an improved color contrast. The average prevalence decreased from 12.53\% in \(v1\) to 3.98\% with \texttt{SK\_con}, with a statistical significance (Wilcoxon signed-rank $p<.001$).

Similarly, accessibility score increased by 4.50 points. Table~\ref{tab:rq2-results-03-v3} average outcomes for \(v1\), \(v2\) and \(v3\) runs. While the improvement was stark in \(v1\)->\(v3\) pairs, the same is not true when \(v2\) and \(v3\) runs are compared. Our results indicate that there is no significant difference in either accessibility score or SC 1.4.3 prevalence between a general Skill and a specialized Skill.

\begin{table*}[t]
  \centering
  \caption{Matched accessibility outcomes after inclusion of $SK_{con}$}
  \label{tab:rq2-results-03-v3}
  \small
  \begin{tabular}{lrrrr}
    \toprule
    Outcome & \(v1\), mean (SD) & \(v2\), mean (SD) & \(v3\), mean (SD) & \(v1\)--\(v3\) change ($p$) \\
    \midrule
    SC~1.4.3 violation rate (\%) & 12.53 (11.07) & 4.60 (7.14) & 3.98 (4.67) & $-8.55$ ($<.001$) \\
    Accessibility score         & 88.06 (6.37)  & 92.13 (5.51) & 92.56 (3.21) & $+4.50$ (.003) \\
    \bottomrule
  \end{tabular}
\end{table*}

\subsubsection{Functional correctness}

\begin{table*}[t]
  \centering
  \caption{Functional correctness across matched no-skill and skill-assigned runs.}
  \label{tab:rq2-functional-correctness}
  \small
  \begin{tabular}{lrrrrrr}
    \toprule
    Project & $n$ & \(v1\) & \(v2\)
      & $\Delta$ \(v2\)--\(v1\) & \(v3\) & $\Delta$ \(v3\)--\(v1\) \\
    \midrule
    $P_1$-Catalog  & 8  & 99.26 & 85.29 & $-13.97$ & 97.06 & $-2.21$ \\
    $P_2$-Planner    & 8  & 93.06 & 62.50 & $-30.56$ & 95.14 & $+2.08$ \\
    $P_3$-Viz  & 8  & 100.00 & 87.50 & $-12.50$ & 90.63 & $-9.38$ \\
    \midrule
    Overall & 24 & 97.44 & 78.43 & $-19.01$ & 94.27 & $-3.17$ \\
    \bottomrule
  \end{tabular}
\end{table*}

We also compared the functional correctness of both Skill runs, and compared with the high scores of \(v1\) runs. We observed a substantial decrease in \(v2\). As shown in Table~\ref{tab:rq2-functional-correctness}, mean correctness fell from 97.44 in \(v1\) to 78.43 in \(v2\), a paired decrease of 19.01 points.
\(v3\) runs showed a smaller and less consistent difference. Mean correctness was 94.27 in \(v3\), 3.17 points below \(v1\).

\subsubsection{Trajectory analysis}

Similar to RQ1, we delved deeper into the website creation runs and analyzed their trajectory characteristics. 

\begin{table*}[t]
  \centering
  \caption{Trajectory activity across the 24 matched configurations.}
  \label{tab:rq3-matched-trajectory-activity}
  \small
  \begin{tabular}{lrrrrrrrr}
    \toprule
    & \multicolumn{4}{c}{Mean semantic actions per run}
    & \multicolumn{2}{c}{Accessibility reasoning}
    & \multicolumn{2}{c}{Accessibility verification} \\
    \cmidrule(lr){2-5}\cmidrule(lr){6-7}\cmidrule(lr){8-9}
    Version & Total & Plan & Implement & Verify
      & Runs & Mean count & Runs & Mean count \\
    \midrule
    \(v1\) & 76.46 & 22.46 & 15.67 & 38.33 & 11/23 & 1.04 & 0/24  & 0.00 \\
    \(v2\) & 91.92 & 22.29 & 15.04 & 54.58 & 22/22 & 10.00 & 11/24 & 1.63 \\
    \(v3\) & 89.25 & 21.00 & 10.50 & 57.75 & 20/22 & 5.59 & 19/24 & 6.92 \\
    \bottomrule
  \end{tabular}
\end{table*}

\begin{figure*}[t]
  \centering
  \includegraphics[width=\textwidth]{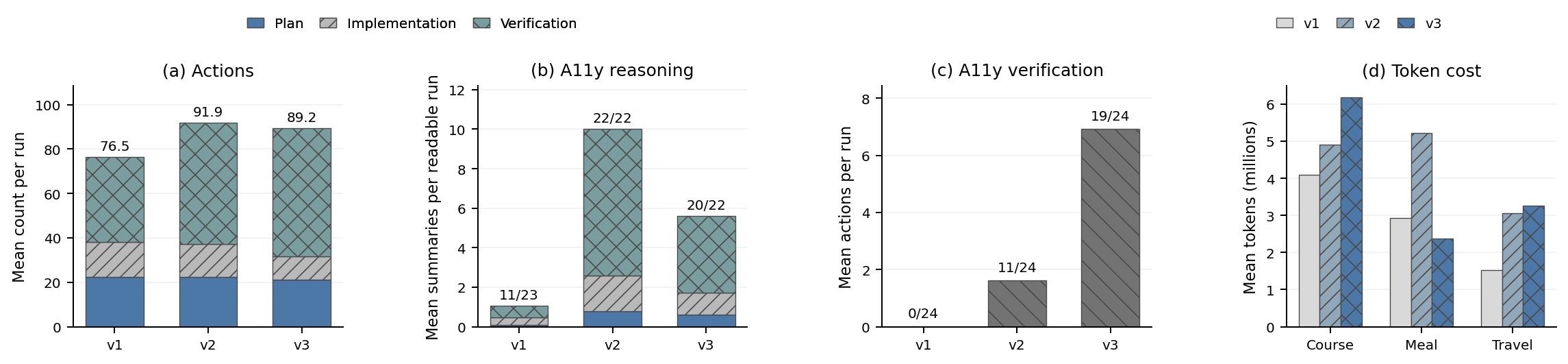}
  \caption{Trajectory changes across matched \(v1\), \(v2\), and \(v3\) configurations.
  From left to right, panels show mean semantic actions by phase (a), mean accessibility-related
  reasoning summaries by phase (b), mean accessibility-verification actions (c), and
  mean total token count by project (d).}
  \Description{There are four bar charts. The first ones shows phase-wise action count for v1, v2 and v3 runs. The second one shows phase-wise accessibility reasoning count for v1, v2 and v3 runs. The third one shows accessibility verification action counts for v1, v2 and v3 runs. Lastly, the fourth one shows token cost comparison between v1, v2 and v3 runs for the three projects individually.}
  \label{fig:rq3-trajectory-changes}
\end{figure*}

Mean action volume increased from 76.46 actions in \(v1\) to 91.92 in \(v2\) and 89.25 in \(v3\) (Table~\ref{tab:rq3-matched-trajectory-activity}). The increase was concentrated in verification, which rose from 38.33 actions in \(v1\) to 54.58 in \(v2\) and 57.75 in \(v3\) (Fig.~\ref{fig:rq3-trajectory-changes}(a)). This can be attributed to the accessibility auditing Skills injected, as we observe a sharp increase in accessibility-related verification actions in \(v2\) (39 actions; mean 1.63), but especially in \(v3\) (166 actions; mean 6.92). The \(v2\) calls mainly ran Axe, Lighthouse, browser audits, keyboard checks, and semantic checks. The \(v3\) calls were more targeted: 157 of 166 executed the provided contrast-checker scripts.

Accessibility reasoning increased from 1.04 summaries per readable \(v1\) run to 10 in \(v2\), then declined to 5.59 in \(v3\). This can be attributed to the textual guidance and multiple WCAG SCs included in $SK_{gen}$, as opposed to scripts on a single SC in $SK_{con}$. The content reflected the intervention: \(v2\) reasoning broadly covered ARIA, keyboard, focus, structure, and automated auditing, whereas \(v3\) reasoning emphasized minimum contrast alongside accessible names and focus states. Its timing also shifted toward verification, as shown in Fig.~\ref{fig:rq3-trajectory-changes}(b).

Mean total token count rose from 2.84 million in \(v1\) to 4.36 million in \(v2\) (+53.4\%) and 4.01 million in \(v3\) (+41.1\%). The project-level breakdown, which differed in token cost, are shown in Fig.~\ref{fig:rq3-trajectory-changes}(d).

\section{Discussion}
\label{sec:discussion}

\subsection{Accessibility Pitfalls in Agent-Generated Websites}

The generated websites exhibited both systematic and component-specific accessibility failures. Minimum color contrast was violated in all 90 baseline runs and affected the largest number of elements. These failures commonly recurred across secondary text, metadata, navigation states, helper text, tables, and buttons, suggesting that a single inaccessible color token or style decision can propagate throughout an otherwise polished interface. Other failures depended on project complexity: catalog websites contained keyboard-inaccessible scrollable regions and accessible-name mismatches; planner websites exposed unlabeled form controls, unnamed buttons, small targets, and malformed ARIA structures; and visualization websites presented charts and SVGs without reliable representations for assistive technology. 

The interventions also revealed that plausible repairs can introduce new barriers. In several general-Skill runs, agents added more descriptive \texttt{aria-label} values but did not preserve the visible wording, increasing label-in-name violations. This may improve context for screen-reader users while creating difficulties for speech-input users. This issue is consistent with Suh et al.'s findings~\cite{suh2026empirical}, where they observed that LLMs inject more ARIA in the code when prompted for accessibility, but increases ARIA-related violations. Guidance should prioritize native semantic HTML, discourage unnecessary ARIA overrides, and require agents to rerun audits after remediation.

\subsection{Implications for Vibe Coding}

Our findings suggest that inaccessibility will remain a default outcome of vibe coding unless accessibility is made part of the agent's completion criteria. Baseline agents produced highly functional websites, yet none produced a fully accessible one and none performed an accessibility-specific verification action. Greater reasoning effort did not significantly improve accessibility, and accessibility-related reasoning alone was not associated with the final score. Thus, neither functional testing nor general model deliberation reliably exposes accessibility failures.

This necessitates accessibility-focused interventions. Prior work have looked into prompting as an intervention~\cite{aljedaani2024does, abu2025can, leedy2025accessibility}. But that requires the developer to be aware of accessibility and change prompts deliberately, which neither the vibe coding process~\cite{chou2026building} nor general awareness of accessibility~\cite{huq2023a11ydev} facilitates. Our experiment therefore examined the result of implicitly invoking Skills during website creation. The findings showed promise, as accessibility scores rose on average, and specific issue like color contrast decreased drastically. We observed how the agents reasoned about accessibility more when exposed to relevant resources, and how accessibility-related actions increased during verification, auditing and fixing emerging issues. The latter supports prior work that looked into iterative-testing as an effective strategy~\cite{aljedaani2024does, aljedaani2026multi, suh2026empirical}. Therefore, if models alone cannot generate accessible websites, accessibility-focused Skills with accessibility guidelines, example code and dynamic auditing scripts can be deployed as a default Skill in coding agents.

However, such specialized interventions must also be associated with appropriate functional testing. Similar to how prior work demonstrated agents' declining functional correctness with the inclusion of specific constraints like style~\cite{zhong2025swe} and security~\cite{zhao2025vibe}, our experiments showed a decrease of 19.01 point in functionality score between no-Skill and general-Skill runs. Additionally, following the same prior work, we saw some introduction of new accessibility issues despite the added constraints. An accessibility intervention should therefore be evaluated not only by the violations it removes, but also by the new accessibility and functional regressions it introduces.

\subsection{Future Directions}

Our work provides empirical evidence of inaccessibility and its remediation with Skills mined from GitHub. Future work can conduct more targeted experiments with specialized Skills and varying resources, to compare effectiveness of textual guidance, code examples, static scripts, dynamic testing and more. Some work have used LLMs to detect accessibility issues~\cite{huq2025automated, he2025enhancing, he2026reflair}, which can be integrated into the website creation process for a more robust intervention. Studies can also expand to other frameworks and platforms, like Android, iOS and Desktop applications. Furthermore, multi-shot development, where subsequent prompts are provided for feature addition or bug fixing~\cite{chou2026building, zhong2026vibench}, needs to be investigated to check how accessibility evolves with each consecutive prompt response. Lastly, Skills in our experiments were designed for auditing and fixing. Future research can look into interventions earlier in the process, during planning and design.

\section{Threats to Validity}
\label{sec:threats}

\textit{Construct Validity.}
Accessibility scores are reliant on automated checkers. We mitigated dependence on any single evaluator by converging results from axe-core, IBM Equal Access Accessibility Checker, and QualWeb, and we complemented those results with testing by blind users.

Functional-correctness uses assertions derived from project rule sets. Manual validation of a subset achieved 97.7\% agreement with automated results, increasing confidence in the measure.

\textit{Internal Validity.}
LLM generation is non-deterministic, so differences between matched runs cannot be attributed exclusively to the Skill text. Our paired comparisons hold project and agent--model--effort configuration constant, and the baseline pool includes replications for core configurations, but random generation variability remains.

Instead of writing our own Skills, which would question the validity of our constructs, we mined GitHub for the most popular and commonly used Skills. However, that does not eliminate the inherent limitations of those Skills. Hence, they should be understood as representative intervention packages rather than controlled frontier artifacts.

\textit{External Validity.}
Our selection of websites and implementation technology do not cover the full diversity of websites, but we aimed to maximize representativeness through a three-point triangulation process. Regardless, the results should not be fully interpreted for other platforms, design systems, or application domains.

The evaluated agents and models represent contemporary proprietary and open alternatives at the time of the study. These systems evolve rapidly and the relative model results should not be assumed to persist across future versions. The more durable finding is that all tested configurations produced violations and that baseline workflows lacked accessibility-specific verification.
\section{Conclusion}
\label{sec:conclusion}

Vibe coding makes website creation faster and more convenient to developers, but it does not yet make the resulting websites reliably accessible to users. Across 90 websites generated by contemporary coding agents, none was free of accessibility violations, despite high functional correctness. Accessibility-focused Skills improved accessibility outcomes, owing to the increased consideration accessibility during its reasoning and verification stages. However, neither produced fully accessible websites and functional correctness regressed. Therefore, for coding agents to support an inclusive future of software creation, agent environments must provide accessibility-focused interventions and not rely solely on the base model and developer awareness.

\section{Data Availability}
All relevant data used and produced in this study are available in the replication package~\cite{replication_package}.

\bibliographystyle{ACM-Reference-Format}
\bibliography{main}

\end{document}